\documentstyle[twocolumn,prl,aps,epsf]{revtex}

\newcommand{\vct}[1]{{\bf #1}}
\newcommand{\vctgk}[1]{\mbox{\boldmath{$#1$}}}
\newcommand{\brkt}[3]{\langle#1|#2|#3\rangle}

\begin{document}

\draft

\title{Effects of the Neutron Spin-Orbit Density on Nuclear
Charge Density in Relativistic Models}

\author{Haruki Kurasawa}
\address{Department of Physics, Faculty of Science, Chiba University,
Chiba 263-8522, Japan}

\author{Toshio Suzuki}
\address{Department of Applied Physics, Fukui University, Fukui 910-8507,
Japan\\
RIKEN, 2-1 Hirosawa, Wako-shi, Saitama 351-0198, Japan}

\maketitle

\begin{abstract}
The neutron spin-orbit density contributes to the nuclear
charge density as a relativistic effect. The contribution is
enhanced by the effective mass stemming from the Lorentz-scalar
potential in relativistic models. This enhancement explains well
the difference between the
cross sections of elastic electron scattering off $^{40}$Ca and
$^{48}$Ca which was not reproduced in non-relativistic models.
The spin-orbit density will be examined in more detail in
electron scattering off unstable nuclei which would be
available in the future. 
\end{abstract}

\pacs{PACS numbers: 21.10.Ft, 25.30.Bf}

At present there are two kinds of phenomenological models which
explain well nuclear structure and reactions. The one is a
conventional non-relativistic model which assumes phenomenological
interactions between nucleons, like Skyrme 
forces\cite{vb}. The other is a relativistic model which takes
into account explicitly meson-exchanges between nucleons,
as the $\sigma-\omega$ model\cite{sw}. The relationship between
these models, however, is not clear. In the relativistic
model the effective mass coming from the Lorentz scalar
potential, which is absent in non-relativistic models, plays a
crucial role in understanding nuclear properties.
For example, the effective mass enhances the spin-orbit force which is
responsible for the spin-orbit splitting of the single-particle
levels and polarization phenomena of hadron-nucleus
scattering. The nuclear saturation density and binding energy
are also dominated by the effective mass. In non-relativistic
models those quantities are explained by taking into account
other kinds of many-body correlations. It is one of the important
questions in nuclear physics which model is realistic.
The best way to answer this question is to
find fundamental physical quantities which can be explained in the
one model, but not in the other model. So far, however, such
quantities are not found, as far as the authors know.

The purpose of this paper is to show that the neutron
spin-orbit density is very sensitive to the effective mass of the
relativistic model. The effective mass yields an additional
spin-orbit density as a relativistic correction, which
contributes to the nuclear charge density, and the effects are
seizable in the form factors of elastic electron scattering.
This correction explains well the existing data which were not
reproduced in  previous non-relativistic models\cite{ber}.
We will also show that the effective mass effects will be
able to be explored in more detail in unstable nuclei.

The neutron spin-orbit charge density is due to the Pauli
current. It was first discussed by Bertozzi {\em et al.}\cite{ber} in a
non-relativistic framework. The nuclear wave function was
obtained in a two-component model and the relativistic
correction was derived by expanding the free nucleon current
in terms of $1/M$, $M$ being the mass of the free nucleon.
They calculated the cross sections for elastic electron
scattering off $^{40}$Ca and $^{48}$Ca, and found that the
relativistic correction was not negligible, but not enough
to explain the difference
between the two cross sections. Later Miller\cite{mi}
analyzed the same data using a relativistic model, but
again could not reproduce the experimental data. The Pauli current
made rather worse the agreement between his results and experiment.

Our model discussed below is in principle the same as Miller's
one, but we will calculate the spin-orbit density, using a
different relativistic model developed later by Horowitz
and Serot\cite{hs}. Moreover we will make clear the relationship
between the relativistic model and non-relativistic models.
The effective mass effects, which are peculiar to the
relativistic model, on the spin-orbit current will be clarified.

We calculate the cross section for elastic electron scattering
using phase shift analyses\cite{ye}. For this purpose we
have to obtain the nuclear charge density, which is given by
\begin{equation}
\rho_c(r) = \int\frac{d^3q}{(2\pi)^3}\exp(-i\vct{q}\cdot\vct{r})
 \brkt{0}{\hat{\rho}(\vct{q})}{0},\label{c}
\end{equation}
where $\vct{q}$ denotes the momentum transfer from the electron
to the nucleus. In the relativistic theory, the ground-state
expectation value of the time-component of the nuclear current
is given by
\begin{eqnarray}
\brkt{0}{\hat{\rho}(\vct{q})}{0}
&=&\langle 0| \sum_k\exp(i\vct{q}\cdot\vct{r}_k) \nonumber \\
& &{}\times \left(
F_{1k}(\vct{q}^2)
+\frac{\mu_k}{2M}F_{2k}(\vct{q}^2)\vct{q}\cdot\vctgk{\gamma}_k
\right)|0\rangle ,
\end{eqnarray}
where $F_{1k}(\vct{q}^2)$ and $F_{2k}(\vct{q}^2)$ stand for
the Dirac and Pauli form factors of the nucleon, respectively,
and $\mu_k$ the anomalous magnetic moment. The above equation
is rewritten by using the Sachs form factor,
$G_E(\vct{q}^2)$, as
\begin{eqnarray}
\brkt{0}{\hat{\rho}(\vct{q})}{0}
&=& \int d^3x\exp(i\vct{q}\cdot\vct{x}) \nonumber \\
& &
\times\sum_\tau\left(G_{E\tau}(\vct{q}^2)\rho_\tau(x)
	  +F_{2\tau}(\vct{q}^2)W_\tau (x)\right)\nonumber\\
  &=& \int d^3x\,d^3y\exp(i\vct{q}\cdot(\vct{x} + \vct{y})) \nonumber \\
& & \times
   \sum_\tau\left(G_{E\tau}(y)\rho_\tau(x)+F_{2\tau}(y)W_\tau(x)\right),
   \label{fc}
\end{eqnarray}
where the sum of $\tau$ is performed with respect to the proton
and the neutron, $\tau = p, n$. 
The functions, $G_{E\tau}(y)$ and $F_{2\tau}(y)$, are obtained by
the inverse Fourier transformation of the Sachs
and Pauli form factors, respectively.
The nucleon density, $\rho_\tau(x)$, and the spin-orbit density,
$W_\tau(x)$, are given by
\begin{eqnarray}
\rho_\tau(r)&=&\brkt{0}{\sum_k\delta(\vct{r}-\vct{r}_k)}{0},\label{bd}\\
W_\tau(r)&=&\frac{\mu_\tau}{2M}
\left(-\frac{1}{2M}\vctgk{\nabla}^2\rho_\tau(r)\right. \nonumber \\
& & \left.
+\,i\vctgk{\nabla}\cdot
\brkt{0}{\sum_k\delta(\vct{r}-\vct{r}_k)\vctgk{\gamma}_k}{0}\right),
\label{sd}
\end{eqnarray}
where the sum over $k$ is performed up to $Z$ for $\tau=p$ and
up to $N$ for $\tau=n$.
By inserting Eq.~(\ref{fc}) into Eq.~(\ref{c}), the nuclear
charge density is given by
\begin{equation}
 \rho_c(r)=\sum_\tau\left(\rho_{c\tau}(r)+W_{c\tau}(r)\right),
\label{full}
\end{equation}
where the nucleon charge density, $\rho_{c\tau}(r)$, and
the spin-orbit charge density, $W_{c\tau}(r)$, are written as
\begin{eqnarray}
\rho_{c\tau}(r)&=&\frac{1}{r}\int_0^\infty\!dx\,x\rho_\tau(x)
 \nonumber \\
& &{}\times\Bigl(g_\tau(|r-x|)-g_\tau(r+x)\Bigr),\\
W_{c\tau}(r)&=&
\frac{1}{r}\int_0^\infty\!dx\,xW_\tau(x) \nonumber \\
& &{}\times \Bigl(f_{2\tau}(|r-x|)-f_{2\tau}(r+x)\Bigr)
\end{eqnarray}
with
\begin{eqnarray*}
 g_\tau(x)&=&\frac{1}{2\pi}\int_{-\infty}^\infty\!dq\,
 e^{iqx}G_{E\tau}(\vct{q}^2)\,,\\
 f_{2\tau}(x)&=&\frac{1}{2\pi}\int_{-\infty}^\infty\!dq\,
 e^{iqx}F_{2\tau}(\vct{q}^2).
\end{eqnarray*}

In order to calculate the nucleon and the spin-orbit density,
we take the Horowitz-Serot model\cite{hs}, where the ground
state wave function is described with the mean field approximation.
The single-particle wave function is written as
\[
 \psi_{\alpha m}
 =\left(
\begin{array}{c}
i\,\displaystyle{\frac{G_\alpha(r)}{r}}\,{\cal Y}_{\ell jm}\\
 \noalign{\vskip6pt}
\displaystyle{
 \frac{F_\alpha(r)}{r}
 \frac{\vctgk{\sigma}\cdot\vct{r}}{r}
 }\,{\cal Y}_{\ell jm}\\
\end{array}
 \right),%%\qquad (\alpha\neq m).
\]
The large and small components in the present model satisfy
the Dirac equation:
\begin{eqnarray}
\frac{dG_\alpha}{dr}&=&-\,\frac{\kappa_\alpha}{r}G_\alpha
 +\Bigl( \varepsilon_\alpha-U_\tau(r)+M^\ast(r)\Bigr)F_\alpha,
 \nonumber\\
\frac{dF_\alpha}{dr}&=&\frac{\kappa_\alpha}{r}F_\alpha
 -\Bigl(\varepsilon_\alpha-U_\tau(r)-M^\ast(r)\Bigr)G_\alpha,
\end{eqnarray}
where $\kappa_\alpha=(-1)^{j-\ell+1/2}(j+1/2)$ denotes the
eigenvalue of ${}-(1+\vctgk{\sigma}\cdot\vctgk{\ell})$,
and $M^*(r)$ the nucleon effective mass given by
\[
M^*(r) = M-U_s(r).
\]
The Lorentz scalar potential, $U_s(r)$, comes from the
$\sigma$-meson exchanges between nucleons, while the Lorentz vector
potential, $U_\tau(r)$, is due to the $\omega$- and $\rho$-mesons
and photons in the present model.
Then the nucleon density in Eq.~(\ref{bd}) is given by
\begin{eqnarray}
 \rho_\tau(r)
=\sum_{\alpha}\frac{2j_\alpha+1}{4\pi r^2}
\left( G_\alpha^2+F_\alpha^2 \right).
\end{eqnarray}
On the other hand, the spin-orbit density, $W_\tau(r)$,
in Eq.~(\ref{sd}) is described as
\begin{eqnarray}
 W_\tau(r)
 &=&\frac{\mu_\tau}{M}\sum_\alpha\frac{2j_\alpha+1}{4\pi r^2}
\frac{d}{dr}\left(
\frac{M-M^\ast(r)}{M}\,G_\alpha F_\alpha \right. \nonumber \\
& & \left. +\,\frac{\kappa_\alpha+1}{2Mr}\,G_\alpha^2
-\frac{\kappa_\alpha-1}{2Mr}\,F_\alpha^2
\right). \label{rsd}
\end{eqnarray}

The relationship between the relativistic model and
non-relativistic models is very clear. In non-relativistic models,
usually the neutron charge and spin-orbit charge densities are
neglected in estimating the electron scattering cross
section. Bertozzi {\em et al.}\cite{ber} took into account the
neutron charge density and a part of the spin-orbit charge
density in the non-relativistic framework. Their spin-orbit
density is obtained from Eq.~(\ref{rsd}) by setting $M^*(r)=M$
and neglecting $F_\alpha^2$-term,
\begin{eqnarray}
 W_\tau(r)
&\approx& \frac{\mu_\tau}{2M^2 r^2}\frac{d}{dr}\,r
 \sum_\alpha\frac{2j_\alpha+1}{4\pi r^2}
\left(\kappa_\alpha+1\right)\,G_\alpha^2 \\
\noalign{\vskip4pt}
&\approx& -\,\frac{1}{r^2}\frac{d}{dr}\,r\,
\brkt{0}{ \frac{\mu_\tau}{2M^2}
\sum_k\delta(\vct{r}-\vct{r}_k)\,\vctgk{\sigma}_k\!\cdot\!
\vctgk{\ell}_k}{0}.\label{nsd}
\end{eqnarray}
We will show that the spin-orbit density due to the effective mass
in Eq.~(\ref{rsd}) is very important in the relativistic model
for reproducing the experimental data.

The nucleon form factors used in the present calculations are the
dipole type according to Ref.~\cite{ber} with more recent experimental
data for the neutron\cite{lu}. The Sachs form factor for the
proton is given by
\[
 G_{Ep}=\frac{1}{\left(1+r_p^2\vct{q}^2/12\right)^2}, \, \quad
 r_p=\langle r^2\rangle^{1/2}=0.81\,\mbox{fm},
\] 
while the one for the neutron by
\begin{eqnarray*}
G_{En}&=&\frac{1}{\left(1+r_+^2\vct{q}^2/12\right)^2}
 -\frac{1}{\left(1+r_-^2\vct{q}^2/12\right)^2}\,,\\
 r_\pm^2&=&(0.9)^2\mp 0.06\,\mbox{fm}^2.
\end{eqnarray*}
The Pauli form factors for the proton and the neutron are
\[
 F_{2p}=\frac{G_{Ep}}{1+\vct{q}^2/4M^2}\,,\qquad
 F_{2n}=\frac{G_{Ep}-G_{En}/\mu_n}{1+\vct{q}^2/4M^2}.
\]
The values of the anomalous magnetic moment are given by
$\mu_p=1.793$ and $\mu_n=-\,1.913$.

%%%%%%%%
\begin{figure}[hbt]
\epsfxsize=\hsize\epsfbox{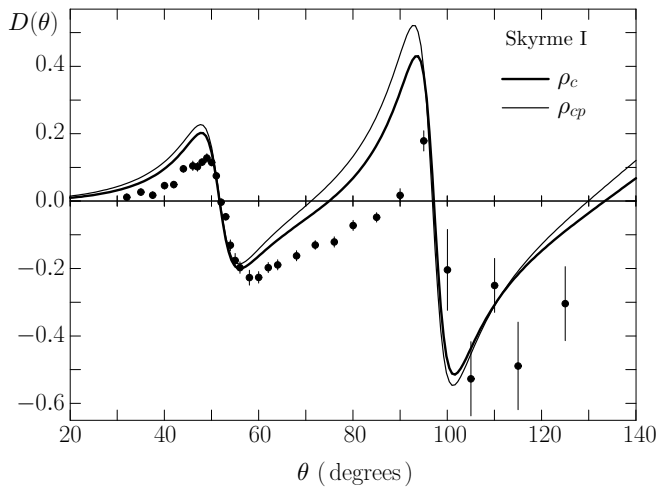}\vspace{12pt}
\caption{The difference $D(\theta)$ for $^{40}$Ca and $^{48}$Ca
given by Eq.~(\protect\ref{dsigma}) as a
function of scattering angle $\theta$ for elastic electron scattering
at 250 MeV. Numerical values are calculated by the non-relativistic
Hartree-Fock approximation with the Skyrme force I.
The thin solid curve shows the results with the only proton charge
density, while the solid curve with the full charge density.
Experimental data points are taken from Ref.\protect\cite{exp}.
}
\end{figure}
%%%%%%%%

Now, we calculate the differential cross sections $\sigma(\theta)$
for elastic electron scattering off $^{40}$Ca and $^{48}$Ca and
compare their difference $D(\theta)$ given by
\begin{equation}
 D(\theta)=\frac{\sigma_{40}(\theta)-\sigma_{48}(\theta)}
 {\sigma_{40}(\theta)+\sigma_{48}(\theta)}, \label{dsigma}
\end{equation}
with experiment. First we show in Fig.~1 results
of the non-relativistic models, since Bertozzi {\em et al.}\cite{ber}
did not compare explicitly their results including both
the nucleon charge density and the spin-orbit charge
density with experiment. In Fig.~1, the thin solid curve shows
the result taking into account the only proton charge
density, which is calculated with the Skyrme force I in
the Hartree-Fock approximation\cite{vb}. This is almost the same
as the result of Bertozzi {\em et al.}\cite{ber}. When we include also
the neutron charge density and the proton and neutron
spin-orbit charge density in Eq.~(\ref{nsd}), we obtain the solid
curve in Fig.~1. It is seen that the the discrepancy
between the theory and the experiment is a little reduced,
in particular, at the electron scattering angle around
$\theta=60^\circ$ to 90$^\circ$. The improvement is
mainly due to the spin-orbit charge density from the neutrons in the
$1f_{7/2}$ shell, but is not enough to explain the experimental
data\cite{exp}.

%%%%%%%%%%%
\begin{figure}[hbt]
\epsfxsize=\hsize\epsfbox{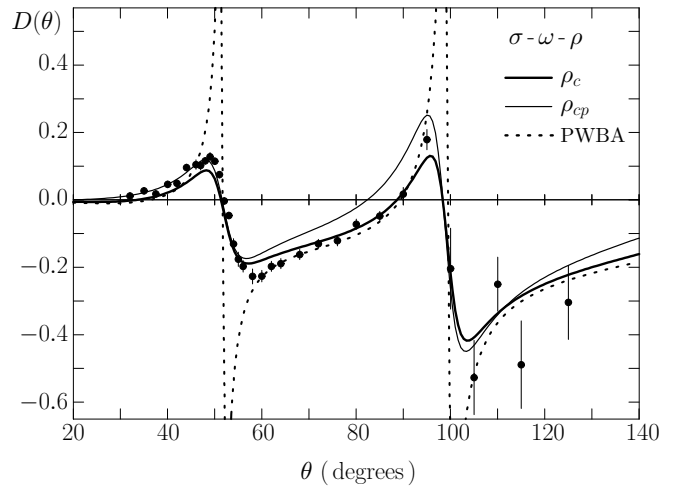}\vspace{12pt}
\caption{Same as Fig.~1, for the numerical results of the
relativistic model. The dotted curve is calculated in PWBA with the
effective momentum transfer.}
\end{figure}
%%%%%%%%%%%

Fig.~2 shows the results of the Horowitz-Serot model\cite{hs}.
The thin solid curve is obtained by taking into account the
only proton charge density, while the solid curve by the full
density, Eq.~(\ref{full}). In the relativistic model, the proton charge
density itself improves results of the non-relativistic model,
and the experimental data are almost reproduced by taking
into account the neutron spin-orbit charge density enhanced by
the effective mass. In this figure, we show, as a
reference, by the dotted curve the results of the PWBA
calculations using the full density and the effective momentum
transfer to simulate DWBA. We note that both in non-relativistic
and relativistic models, the center of mass correction to
the cross section are negligible in Ca isotopes\cite{kms}.

As seen in Eq.~(\ref{rsd}), the effects of the spin-orbit charge
density appears, when the sub-shell is occupied by the
neutrons; In closed shell nuclei, the effects disappear.
Moreover, if protons also occupy the subshell as in $^{208}$Pb,
the proton and neutron spin-orbit densities almost cancel each
other as in non-relativistic model\cite{ber}, since the
anomalous magnetic moment of the proton has the opposite
sign to that of the neutron. Another interesting result of the
spin-orbit density is found in neutron rich nuclei.
In Fig.~3 shown the results with respect to $^{40}$Ca and $^{52}$Ca in
the same designation as in Fig.~2. We see that effects from
the spin-orbit charge density of the $1f_{7/2}$ neutrons are
almost cancelled by those from the $2p_{3/2}$ neutrons.
The similar results are obtained in Zr isotopes\cite{kms}. The effect
of the neutron spin-orbit charge density is enhanced in the cross
section of $^{90}$Zr, compared with the one of
$^{80}$Zr, but disappears in $^{96}$Zr. It is interesting to
observe experimentally these predictions of the relativistic
model in electron scattering off unstable nuclei which is
planned in RIKEN\cite{suda}.

%%%%%%%%%%%
\begin{figure}[hbt]
\epsfxsize=\hsize\epsfbox{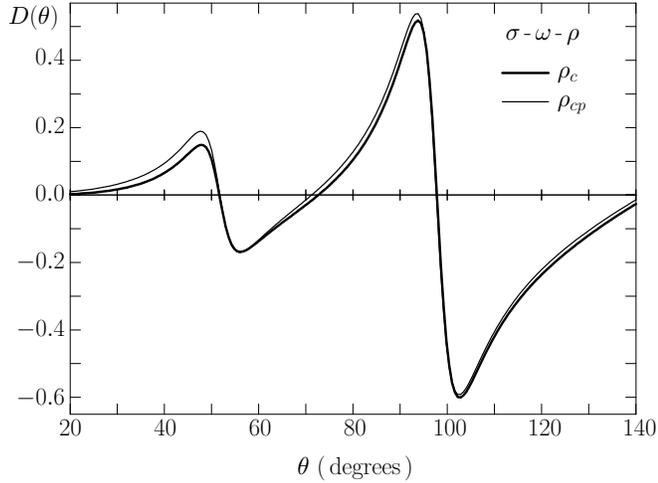}\vspace{12pt}
\caption{Same as Fig.~2, for the difference as to $^{40}$Ca and
$^{52}$Ca calculated by the relativistic model.}
\end{figure}
%%%%%%%%%%%%

In conclusion, the effective mass due to the Lorentz scalar
potential is a necessary ingredient of the relativistic model.
It enhances the neutron spin-orbit charge density in a peculiar
way to this model. The enhanced density explains well the
difference between the cross sections of elastic electron
scattering off $^{40}$Ca and $^{48}$Ca, which was not reproduced
in previous non-relativistic models. Electron scattering off
unstable nuclei is desirable in order to explore in more
detail the effective mass effects. More detailed discussions on the
results of non-relativistic and relativistic models
will be published elsewhere\cite{kms}.

\acknowledgments

The authors would like to thank Dr. T. Suda for useful discussions.


\begin{thebibliography}{99}
\bibitem{vb}D.Vautherin and D. M. Brink,
Phys. Rev. {\bf C5} (1972) 626.
\bibitem{sw}B. D. Serot and J. D. Walecka,
Adv. Nucl. Phys. {\bf 16} (1986) 1.
\bibitem{ber}W. Bertozzi, J. Friar, J. Heisenberg and J. W. Negele,
Phys. Lett. {\bf 41B} (1972) 408. 
\bibitem{mi}L. D. Miller, Phys. Rev. {\bf C14} (1976) 706.
\bibitem{hs}C. J. Horowitz and B. D. Serot,
Nucl. Phys. {\bf A368} (1981) 503.
\bibitem{ye}D. R. Yennie, D. G. Ravenhall and R. N. Wilson,
Phys. Rev. {\bf 95} (1954) 500.
\bibitem{lu}%D. H. Lu et al., Phys. Rev. {\bf C57} (1998) 2628.
T. Eden {\em et al.}, Phys. Rev. {\bf C50} (1994) 1749;
M. Meyerhoff {\em et al.}, Phys. Lett. {\bf 327B} (1994) 201;
S. Platchkov {\em et al.}, Nucl. Phys. {\bf A510} (1990) 740.
\bibitem{exp}R. F. Frosch {\em et al.}, Phys. Rev. {\bf 174} (1968) 1380.
\bibitem{kms}H. Kurasawa, H. Madokoro and T. Suzuki, to be published.
\bibitem{suda}T. Suda, private communication.
\end{thebibliography}
\end{document}